\documentclass[twocolumn,english,aps,prl,showpacs,superscriptaddress,preprintnumbers]{revtex4-1}
\usepackage[T1]{fontenc}
\usepackage[latin9]{inputenc}
\usepackage{float}
\usepackage{amsmath}
\usepackage{amssymb}
\usepackage{graphicx}
\usepackage{esint}

\makeatletter
\@ifundefined{textcolor}{}
{%
 \definecolor{BLACK}{gray}{0}
 \definecolor{WHITE}{gray}{1}
 \definecolor{RED}{rgb}{1,0,0}
 \definecolor{GREEN}{rgb}{0,1,0}
 \definecolor{BLUE}{rgb}{0,0,1}
 \definecolor{CYAN}{cmyk}{1,0,0,0}
 \definecolor{MAGENTA}{cmyk}{0,1,0,0}
 \definecolor{YELLOW}{cmyk}{0,0,1,0}
}

\usepackage{amstext}
\usepackage{babel}
\newcommand{\unit}{1\!\!1}
\addto\captionsenglish{}
\makeatother

\begin{document}
\title{Dissipation-driven entanglement between qubits mediated by plasmonic nanoantennas}
\author{J. Hou}
\author{K. S\l owik}
\author{F. Lederer}
\affiliation{Institute of Condensed Matter Theory and Solid State Optics, Abbe
Center of Photonics, Friedrich-Schiller-Universit\"{a}t Jena, D-07743
Jena, Germany}
\author{C. Rockstuhl}
\affiliation{Institute of Theoretical Solid State Physics, Karlsruhe Institute of Technology, 76131 Karlsruhe, Germany}
\affiliation{Institute of Nanotechnology, Karlsruhe Institute of Technology, 76131 Karlsruhe, Germany}

\begin{abstract}
A novel scheme is proposed to generate a maximally entangled state between two qubits by means of a dissipation-driven process. To this end, we entangle the quantum states of qubits that are mutually coupled by a plasmonic nanoantenna. Upon enforcing a weak spectral asymmetry in the properties of the qubits, the steady-state probability to obtain a maximally entangled, subradiant state approaches unity. This occurs despite the high losses associated to the plasmonic nanoantenna that are usually considered as being detrimental. The entanglement scheme is shown to be quite robust against variations in the transition frequencies of the quantum dots and deviations in their prescribed position with respect to the nanoantenna. Our work paves the way for novel applications in the field of quantum computation in highly integrated optical circuits.
\end{abstract}


\maketitle
\section{Introduction\label{sec:introduction}}
The remarkable performance of quantum protocols relies on the entanglement between quantum bits (qubits). Ideally, this entanglement should be preserved over long time scales. Common experience predicts that dissipation and the related decoherence in the system constitute a major obstacle in preserving this entanglement \cite{Chen2012,Jin2013}. However, this intuition has been proven wrong in numerous recent works that focused on the generation of entanglement by coupling qubits to a common, dissipative environment \cite{Ficek2002,Lin2013,Shankar2013,Lee2013}. Such dissipation driven entanglement schemes are extremely appealing from a practical point of view since they allow to achieve entanglement independent of the initial state of the system and were shown to be quite robust against variations in the control parameter.

In general, such dissipative environment can be provided by metallic nanostructures. They are extremely appealing in cavity quantum electrodynamics for developing highly integrated quantum optical circuits where quantum information is generated, processed, and detected at the nanoscale.  At optical frequencies the properties of these metallic nanostructures are dominated by the excitation of surface plasmon polaritons, which are carrier oscillations in the metal resonantly coupled  to an external electromagnetic field. The excitation of surface plasmon polaritons permits the localization and enhancement of electromagnetic fields at length scales adapted to the size of quantum emitters. Simultaneously, the excitation is accompanied by non-radiative dissipation of electromagnetic energy into heat and eventually also by radiative losses. However, this feature, usually considered as a disadvantage,  can be turned into a benefit in dissipation driven processes, where the generation of entanglement ia a prime example.
\begin{figure}[t]
\includegraphics{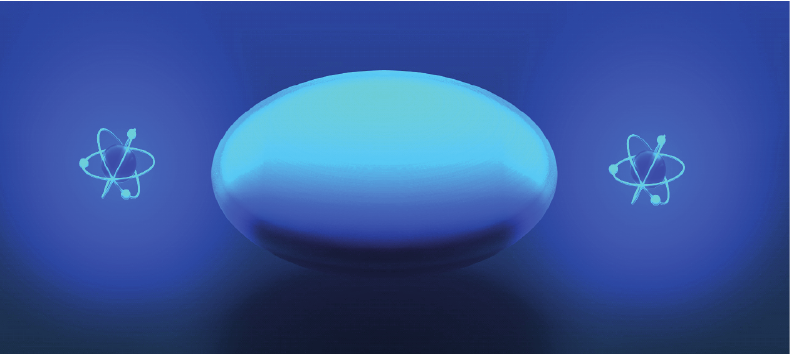}
\caption {\label{fig:scheme} Scheme of the investigated system. Two qubit carriers (eventually atoms, molecules or quantum dots) are positioned in the vicinity of an optical nanoantenna. The system is subject to an external driving field.}
\end{figure}

So far, most of the entanglement-generation protocols and entangling gates have been exploiting nanowires as the specific metallic nanostructure. These metallic nanowires sustain propagating surface plasmon polaritons. Such excitations are surface waves that propagate along the nanowires over extended distances \cite{Barnes2003,Zayats2005,Kewes2013}. In such geometries, the quantum states of qubits encoded in quantum dots, atoms, or even molecules can be entangled by bringing them in close proximity to the nanowire. For identical qubits, a maximal degree of entanglement can be obtained by choosing a suitable relative coupling strengths to the nanowire \cite{Chen2011,Yang2013}. In this scheme, multipartite entanglement between the qubits and the plasmons is achieved and followed by a measurement of a number of plasmonic excitations at the terminations of the nanowire. Only if no plasmons are detected, it can be concluded that the qubits were maximally entangled with each other.

To circumvent this probabilistic character and to generate high degrees of entanglement in deterministic schemes, it was suggested to asymmetrically place the qubits with respect to the nanowire. For definite coupling strengths a maximal degree of entanglement could be achieved regardless of the state of the plasmonic field \cite{Chen2012,Jin2013}. However, large losses associated with propagating surface plasmons prevent this entanglement to persist for longer time scales. An idea of using the loss channels to drive the qubits into a subradiant state that is robust against dissipation was investigated in Ref.~\onlinecite{MartinCano2011}. The described scenario allows for achieving stationary entanglement. However, it requires the system to be initially driven into a predefined state and, moreover, the resulting entanglement remains significantly below the maximal possible degree. Recently, also a scheme for robust-to-loss and high-fidelity entanglement generation was proposed with a waveguide made from a chain of metallic nanoparticles \cite{Lee2013}.

To date, however, none of these schemes could provide a maximal stationary entanglement of a pair of qubits in a deterministic manner. In our study, we report that strongly dissipative systems can be used to achieve this goal. Contrary to much of the previous work that considered propagating surface plasmon polaritons in metallic waveguides, we consider here an isolated metallic nanoantenna as the dissipative structure. The optical nanoantenna sustains localized surface plasmon polaritons that require a proper quantization while studying the coupling to multiple qubits to be entangled \cite{McEnery2013,Artuso2013}. The crucial requirement for entanglement is the existence of a certain degree of asymmetry in the qubit configuration. This asymmetry can be easily enforced by various means. But if the asymmetry is provided by different transition frequencies in the qubit carriers, the resulting entanglement turns out to be especially robust against dissipation and decoherence, and persists even in the steady state. Moreover, our protocol provides stationary entanglement, i.e. it does not depend on the initial state of the system.

Recently, a similar scheme has been proposed for a deterministic multi-qubit quantum phase gate for qubits interacting with suface plasmons of a nanosphere \cite{Ren2014}. Moreover, possibilities of using nanoantennas as entangling devices for photons \cite{Oka2013} and nanoantennas coupled to quantum dots as sources of entangled photon pairs \cite{Hecht2009,Maksymov2012} have recently been discussed.

This paper is organized as follows. In Sect. \ref{sec:model}, we introduce the Tavis-Cummings model with damping to describe the coupling of a pair of qubits to a lossy single-mode nanoantenna field. Next, in Sect. \ref{sec:adiabatic_elimination}) we introduce an effective picture for describing all observables that is valid in the regime of weak qubit-to-nanoantenna coupling. Such a simplified view provides us with an intuitive understanding of the entanglement-generation mechanisms. We verify the expectations based on the effective picture by a fully numerical solution of the evolution of the hybrid quantum system in Sect. \ref{sec:results}. Finally, we analyze the effect of disorder, introduced by, e.g., experimental imperfections that may be present in the proposed scheme, on the degree of entanglement. It is shown that the present scheme is extremely robust; rendering it an ideal candidate for future experiments in the field of quantum plasmonics.

\section{Model\label{sec:model}}
In this section we describe the system to be investigated in more detail. We briefly introduce the Lindblad-Kossakowski formalism that allows in particular to find the steady-state density matrix of the two qubits, and to estimate the corresponding degree of entanglement.

The system consists of two qubits in the vicinity of a metallic nanoantenna, as shown in Fig.~\ref{fig:scheme}. The ground ($|g^{(j)}\rangle$) and excited ($|e^{(j)}\rangle$) states of the $j^\mathrm{th}$ qubit are separated by an energy $\hbar\omega^{(j)}$. Each of the qubits corresponds to a pair of flip operators $\sigma_-^{({j})}=|g^{({j})}\rangle\langle e^{({j})}|$, $\sigma_+^{({j})} = {\sigma_-^{({j})}}^\dagger$ and inversion operator $\sigma_z^{({j})} = |e^{({j})}\rangle \langle e^{({j})}|- |g^{({j})}\rangle \langle g^{({j})}|$.

The scattering and absorption spectra of the adjacent nanoantenna are assumed to be well-characterized by single Lorentzian lines centered at frequency $\omega_\mathrm{na}$. For the parameter range, we are interested in, this holds for the ellipsoidal nanoparticles we consider here as the optical nanoantenna \cite{Slowik2013}. It is only required that the ellipsoids are sufficiently small such that their optical response can be described while considering only an electric dipole moment. The resonance wavelength of the ellipsoid can then be tuned by tailoring the axis ratio \cite{Liaw2005}. Such nanoparticles can readily be provided by chemical means and constitute the base to built more complicated functional plasmonic nanostructures \cite{Muehlig2013}.

With such a geometry for the optical nanoantenna in mind, we describe the field localized by the nanoantenna as a single-mode harmonic oscillator, with the annihilation operator $a$. The widths of plasmonic resonances in the scattering and absorption spectra are given by $\Gamma_\mathrm{sca}$ and $\Gamma_\mathrm{abs}$, respectively. For an effective coupling between the nanoantenna and the qubits, the frequencies $\omega^{({j})}$ must be close to $\omega_\mathrm{na}$, i.e. $|\omega^{({j})}-\omega_\mathrm{na}|\ll \Gamma$, where $\Gamma = \Gamma_\mathrm{abs}+\Gamma_\mathrm{sca}$ is the total energy dissipation rate of the nanoantenna. A monochromatic external driving field of frequency $\omega_\mathrm{dr}$, assumed to be almost resonant with the nanoantenna field, provides the energy to the system. We only take into account the coupling $\Omega$ between the driving field and the nanoantenna, since in many practical cases the direct coupling to the qubits is much smaller and can be neglected.

It is convenient to write the Tavis-Cummings Hamiltonian of the above-described system in the frame rotating with the driving field frequency \cite{Bogoliubov1996,Man2009}:
\begin{eqnarray}
H &=& \sum_{{j}=1,2.}\Delta\omega^{{(j)}}\sigma_+^{({j})}\sigma_-^{({j})}+\Delta\omega_\mathrm{na} a^{\dagger}a \label{eq:hamiltonian}\\
&& -\sum_{{j}=1,2.} g^{({j})}\left(\sigma_+^{({j})}a+a^\dagger \sigma_-^{({j})} \right) -\Omega\left( a+ a^\dagger \right), \nonumber
\end{eqnarray}
where $\Delta\omega^{{(j)}} = \omega^{{(j)}}-\omega_\mathrm{dr}$ and $\Delta\omega_\mathrm{na} = \omega_\mathrm{na}-\omega_\mathrm{dr}$ are the detunings of the driving field from the qubit transition frequencies and the nanoantenna resonance, respectively. In the above Hamiltonian, the first two terms correspond to the free evolution of the qubits and the nanoantenna field, and the latter two describe the qubit-nanoantenna and nanoantenna-drive coupling. In both coupling terms we have applied the rotating wave approximation, and the coupling constants are taken real for simplicity. We have also set the Planck's constant $\hbar=1$ in the whole manuscript.

There are several energy dissipation channels in the considered system. Besides the dominating mechanisms where energy is scattered and absorbed by the nanoantenna at a total rate of $\Gamma$, we take into account dissipation effects induced in the qubit-carriers by photonic vacuum, i.e., the spontaneous emission rate $\gamma^{{(j)}}$ of each qubit. For simplicity, we do not include additional dephasing channels for qubits, which is an approximation commonly made in related literature \cite{MartinCano2011,Zubairy2011,Chen2011,Chen2012,Jin2013,Lee2013,SanchezSoto2013}.

The Lindblad operator reads thus \cite{Kossakowski1972}:
\begin{equation}
\mathcal{L}\left(\rho\right) = D_\Gamma\left(\rho,a,a^\dagger\right)+\sum_{{j}=1,2.}D_{\gamma^{({j})}}\left(\rho,\sigma_-^{({j})},\sigma_+^{({j})}\right),\label{eq:lindblad}
\end{equation}
where $\rho$ is the density matrix of the full system (the qubits and the nanoantenna field) and
\begin{equation}
D_\gamma\left(\rho,A,B\right) = \frac{\gamma}{2}\left(2A\rho B-AB\rho-\rho AB\right).
\end{equation}
Typically, the dissipation rates in atoms or quantum dots (in the range of $1\sim 100$ MHz) are several orders of magnitude smaller than those associated with nanoantennas ($\sim 100$ THz) \cite{Govorov2006}. Nevertheless, they might influence both the results at long time-scales and the steady-state of the system, which will be our main subject of interest.

In principle, another dissipation channel should be taken into account for short interqubit distances, below the wavelength of the qubit transition $2\pi c/\omega^{({j})}$, where $c$ is the vacuum speed of light. The channel arises due to the interaction of the two qubits through the surrounding photonic vacuum. This leads to collective effects such as vacuum-induced sub- and superradiance \cite{Dicke1954,Ficek2002,Garraway2011,Nahmad2013}.
As it will turn out later, these effects can only support the preservation of entanglement in our scheme. We thus do not include them here and consider instead something like the worst-case scenario for the entanglement generation.

The dynamics of the system is described by the Lindblad-Kossakowski equation:
\begin{equation}
\dot{\rho}=-i[H,\rho]+\mathcal{L}\left(\rho\right). \label{eq:lindblad}
\end{equation}
To find the steady-state of the system $\rho (t\rightarrow\infty)$, we solve the above equation with the left-hand side set to zero, using a freely-available quantum optics toolbox \cite{QOtoolbox}. For this purpose we truncate the field's Hilbert space sufficiently large, depending on the driving field strength $\Omega$. The steady-state density matrix of the qubit subsystem $\rho^\mathrm{qb}(t\rightarrow\infty)$ can be found by performing a partial trace over the field degrees of freedom: $\rho^\mathrm{qb}(t\rightarrow\infty) = \mathrm{Tr}_\mathrm{field}\left[\rho(t\rightarrow\infty)\right]$.

Coupling of the qubits to a common electromagnetic mode may lead to their significant entanglement, which can be quantified in terms of concurrence, defined as \cite{Wootters2001}
\begin{equation}
C(\rho^\mathrm{qb})=\mathrm{max} \{0,\lambda_1 - \lambda_2 - \lambda_3 - \lambda_4\},
\end{equation}
where $\lambda_k$ are square roots of eigenvalues of $\rho^\mathrm{qb} \tilde{\rho}^\mathrm{qb}$ in descending order, $\tilde{\rho}^\mathrm{qb}=\left(\sigma_y\otimes\sigma_y\right){\rho^\mathrm{qb}}^{\ast}\left(\sigma_y\otimes\sigma_y\right)$, \mbox{$\sigma_y = \left( \begin{array}{cc}0&-\mathrm{i}\\\mathrm{i}&0 \end{array}\right)$} stands for the Pauli matrix and the asterisk denotes  complex conjugation. The concurrence ranges between $0$ (for product states) and $1$ (for maximally entangled states such as Bell states \cite{Nielsen2000}).  We mentionthat the concurrence and other entanglement measures have been analyzed in the context of plasmonics, e.g. in Ref.~\onlinecite{SanchezSoto2013}.

Before we proceed with the results obtained within the model presented in this section, we will introduce an effective description of the qubit-qubit system, which can be obtained by an adiabatic elimination of the field. Such procedure is strictly valid in the weak qubit-nanoantenna coupling regime, defined as $|g^{({j})}|\ll \Gamma$. As we will see, however, this approximative approach provides us with intuitions that hold also beyond the weak-coupling regime.

\section{Effective dynamics in the weak qubit-nanoantenna coupling regime\label{sec:adiabatic_elimination}}
To find the effective description of the qubit-qubit subsystem we are going to start by analyzing the Heisenberg equations of motion of the qubit and field operators. If the coupling of the field to the qubits is weak compared to the dissipation rate of the nanoantenna, it is possible to eliminate the field degree of freedom by substituting the adiabatic solution to the field. As the result, equations of motion for the qubit operators only are obtained. These three steps will be described in subsection \ref{subsec:field_elimination}. An analysis of the resulting equations will allow to find an effective Hamiltonian and effective Lindblad terms in subsection \ref{subsec:eff_Hamiltonian}. Such procedure leads to a drastic simplification of the Lindblad-Kossakowski formalism. The validity of the resulting simplified approach will be verified in subsection \ref{subsec:validity}. To understand the entanglement-generation mechanisms, we will transform the effective Hamiltonian and Lindblad term to the Dicke basis, more suitable to describe an effectively hybridized system of two qubits (subsection \ref{subsec:Dicke}).
Additionally, an expression for concurrence in terms of only a few density matrix elements will be given. This will allow for a more intuitive analysis of entanglement generation in our scheme in a summarizing subsection \ref{subsec:discussion}.

\subsection{Adiabatic elimination of the field operators \label{subsec:field_elimination}}
The Heisenberg equations of motion of the qubits and the field read:
\begin{eqnarray}
\dot{\sigma}_z^{({j})} &=& -\left(\sigma_z^{\mathrm(j)}+\unit\right)\gamma^{({j})}\label{eq:evol_sigz} \\
&& -2\mathrm{i}g^{({j})}\left( a^\dagger\sigma_-^{({j})}-\sigma_+^{({j})}a\right) +f_z^{({j})}, \nonumber \\
\dot{\sigma}_-^{({j})} &=& -\left(\mathrm{i}\Delta\omega^{({j})}+\frac{\gamma^{({j})}}{2}\right)\sigma_-^{({j})}-\mathrm{i}g^{({j})}\sigma_z^{({j})} a+f_-^{({j})}, \label{eq:evol_sig-}\\
\dot{a} &=& -\left(\mathrm{i}\Delta\omega_\mathrm{na}+\Gamma/2\right)a \label{eq:evol_a}\\
&&+\mathrm{i}\left(\Omega+\sum_{{j}=1,2.}g^{({j})}\sigma_-^{({j})}\right)+f_a, \nonumber
\end{eqnarray}
where $f_z^{(j)}$, $f_-^{(j)}$ and $f_a$ are fluctuation operators included to preserve the commutation relations \cite{Scully1997}.

As we have mentioned in the previous section, the dissipation channel of the nanoantenna field dominates over the one of the qubits by orders of magnitude: $\Gamma \gg \gamma^{({j})}$. In this case, we can adiabatically eliminate the field operators \cite{Messiah1999}. Such elimination consists in substituting the adiabatic expression for the annihilation operator (see also Refs.~\onlinecite{Waks2010,Filter2013})
\begin{equation}
a = \frac{\mathrm{i}\left(\Omega+\sum_{{j}=1,2.} g^{({j})}\sigma_-^{({j})}\right)+f_a}{\mathrm{i}\Delta\omega_\mathrm{na}+\Gamma/2} \label{eq:a_adiabatic}
\end{equation}
into equations (\ref{eq:evol_sigz}) and (\ref{eq:evol_sig-}). We arrive at effective evolution equations:
\begin{eqnarray}
\dot{\sigma}_z^{({j})}(t)&=&-\left(\sigma_z^{({j})}+\unit\right)\gamma^{({j})}_\mathrm{eff} +2i\left(\Omega^{({j})}_\mathrm{eff}\sigma_+^{({j})}-{\Omega^{({j})}_\mathrm{eff}}^\ast \sigma_-^{({j})}\right)\nonumber\\
&&+2\mathrm{i}\xi_\mathrm{eff}\left(\sigma_+^{({j})}\sigma_-^{({k})}-\sigma_+^{({k})}\sigma_-^{({j})}\right)\nonumber \\
&&-\gamma^\prime_\mathrm{eff}\left(\sigma_+^{({j})}\sigma_-^{({k})}+\sigma_+^{({k})}\sigma_-^{({j})} \right)+F_z^{({j})}, \label{eq:eff_sigz}\\
\dot{\sigma}_-^{({j})}(t) &=& -\left(\mathrm{i}\Delta\omega^{({j})}_\mathrm{eff}+\frac{\gamma^{({j})}_\mathrm{eff}}{2}\right)\sigma_-^{({j})}-i\Omega^{({j})}_\mathrm{eff}\sigma_z^{({j})}\nonumber \\
&&+\left(\mathrm{i}\xi_\mathrm{eff}+\frac{\gamma^\prime_\mathrm{eff}}{2}\right)\sigma_z^{({j})}\sigma_-^{({k})}+F_-^{({j})},\label{eq:eff_sig-}
\end{eqnarray}
where $k\neq j$, $\Delta\omega^{({j})}_\mathrm{eff} = \omega^{({j})}_\mathrm{eff}-\omega_\mathrm{dr}$ and we have introduced the single-qubit parameters
\begin{eqnarray}
\gamma^{({j})}_\mathrm{eff} &=& \gamma^{({j})} + \Gamma {g^{({j})}}^2/Z,\label{eq:eff_param_single_qubit}\\
\omega^{({j})}_\mathrm{eff} &=& \omega^{({j})}-\Delta\omega_\mathrm{na}{g^{({j})}}^2/Z,\nonumber\\
\Omega^{({j})}_\mathrm{eff} &=& g^{({j})}\Omega \left( \Delta\omega_\mathrm{na}+\mathrm{i}\Gamma/2\right)/Z,\nonumber\\
Z &=& (\Gamma/2)^2+\Delta\omega_\mathrm{na}^2,\nonumber
\end{eqnarray}
the parameters describing an effective collective behaviour
\begin{eqnarray}
\gamma_\mathrm{eff}^\prime &=& \Gamma g^{(1)} g^{(2)}/Z,\label{eq:eff_param_collective}\\
\xi_\mathrm{eff} &=& -g^{(1)}g^{(2)}\Delta\omega_\mathrm{na}/Z,\nonumber
\end{eqnarray}
and modified fluctuation operators
\begin{eqnarray}
F_z^{({j})} &=&f_z^{({j})}+\frac{2\mathrm{i}g^{({j})}}{\Gamma/2+\mathrm{i}\Delta\omega_\mathrm{na}}\sigma_+^{({j})}f_a+\mathrm{H.c.} ,\nonumber\\
F_-^{({j})} &=&f_-^{({j})}-\frac{\mathrm{i}g^{({j})}}{\Gamma/2+\mathrm{i}\Delta\omega_\mathrm{na}}\sigma_z^{({j})}f_a, \nonumber
\end{eqnarray}
where $\mathrm{H.c.}$ stands for a Hermitian conjugate.

\subsection{Effective Hamiltonian and Lindblad term \label{subsec:eff_Hamiltonian}}
Equations equivalent to (\ref{eq:eff_sigz}) and (\ref{eq:eff_sig-}) can be obtained in the density-matrix approach from an effective Lindblad-Kossakowski equation
\begin{equation}
\dot{\rho}^\mathrm{qb}=-i [H_\mathrm{eff},\rho^\mathrm{qb}]+\mathcal{L}_\mathrm{eff}(\rho^\mathrm{qb}), \label{eq:lindblad_reduced}
\end{equation}
with the effective Hamiltonian (see also \cite{Dzsotjan2011})
\begin{eqnarray}
H_\mathrm{eff} &=& \sum_{{j}=1,2.}\left[{\Delta\omega^{({j})}}_\mathrm{eff}\sigma_+^{({j})}\sigma_-^{({j})}-\left(\Omega^{({j})}_\mathrm{eff}\sigma_+^{({j})} +{\Omega^{({j})}_\mathrm{eff}}^\ast\sigma_-^{({j})}\right)\right]\nonumber \\
&&+\xi_\mathrm{eff}\left(\sigma_+^{(1)}\sigma_-^{(2)}+\sigma_+^{(2)}\sigma_-^{(1)}\right), \label{eq:eff_Hamiltonian}
\end{eqnarray}
and the effective Lindblad term
\begin{eqnarray}
\mathcal{L}_\mathrm{eff}\left(\rho^\mathrm{qb}\right) &=& \sum_{{j}=1,2.}D_{{\gamma}_\mathrm{eff}^{({j})}}\left(\rho^\mathrm{qb},\sigma_-^{({j})},\sigma_+^{({j})}\right)\label{eq:eff_lindblad} \\
&& + \sum_{{k}\neq {j}=1,2.} D_{\gamma_\mathrm{eff}^\prime}\left(\rho^\mathrm{qb},\sigma_-^{({k})},\sigma_+^{({j})}\right). \nonumber
\end{eqnarray}
The obtained Hamiltonian and Lindblad term describe effectively the dynamics of two qubits.
The effects induced by the nanoantenna mode are now described by effective parameters, that can be clearly interpreted:
\begin{itemize}
\item {The interaction to the nanoantenna mode shifts the transition frequency of the $j^\mathrm{th}$ qubit to $\omega^{({j})}_\mathrm{eff}$.}
\item {Each of the qubits is coupled to an external field, whose Rabi frequency has been rescaled by the presence of the nanoantenna to an effective value $\Omega^{({j})}_\mathrm{eff}$.}
\item {Now the nanoantenna mediates an effective dipole-dipole interaction $\xi_\mathrm{eff}$ between the qubits.}
\item {The lifetime of the excited state of the $j^\mathrm{th}$ qubit is significantly shortened, such that the corresponding decay rate is enhanced to the value of $\gamma^{({j})}_\mathrm{eff}$.}
\item {Each of the qubits is capable of modifying the electromagnetic environment of the other. This leads to a collective decay rate $\gamma^\prime_\mathrm{eff}$ and corresponding sub- and superradiance effects \cite{Dicke1954,Garraway2011,Nahmad2013}.}
\end{itemize}
It is worth to stress that the direct interplay between the qubits given by $\xi_\mathrm{eff}$, as well as the collective behaviour described by $\gamma^\prime_\mathrm{eff}$, are qualitatively new features, that have been missing in the generic Hamiltonian (\ref{eq:hamiltonian}) and Lindblad term (\ref{eq:lindblad}).

The above-described effective formalism can be easily generalized to a greater number of qubits. Note also that a similar result for qubits coupled to nanowires was recently derived within the Green's function formalism in Ref.~\onlinecite{Dzsotjan2011}.

\subsection{Validity of the effective description \label{subsec:validity}}
Now the validity of the simplified description will be verified by rigorous full-Hamiltonian calculations.
\begin{figure}
\includegraphics{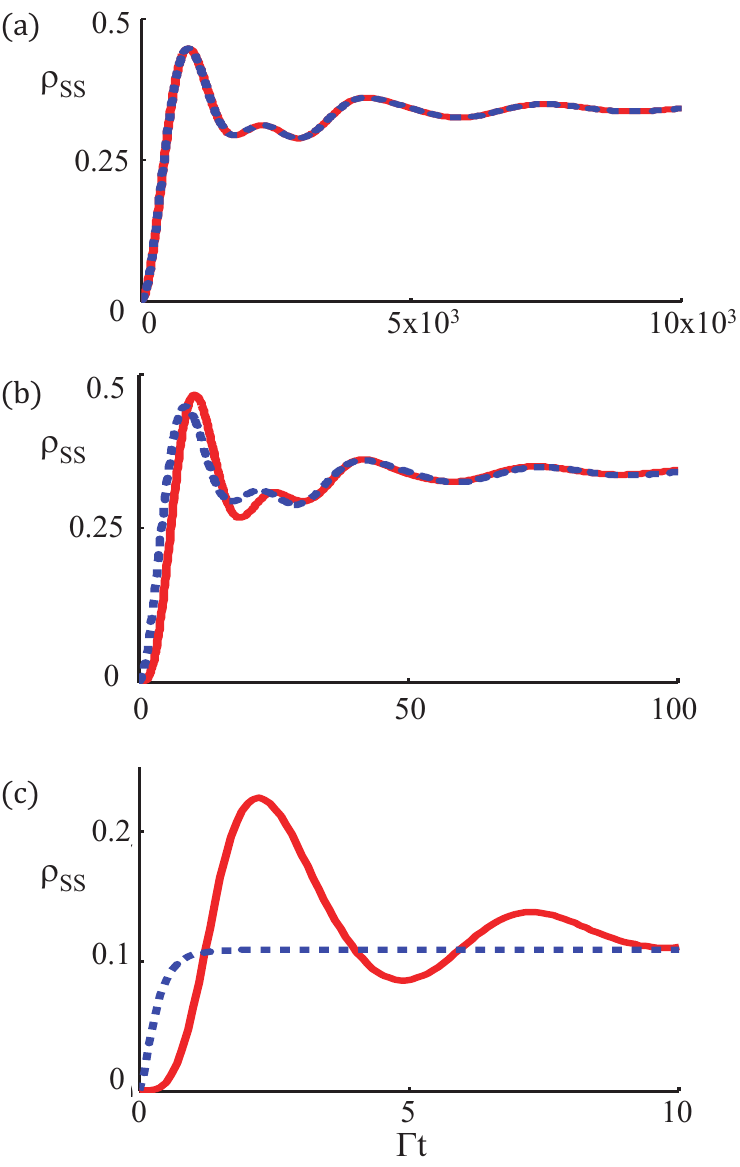}
\caption {\label{fig:adiabatic_approx} Evolution of the symmetric-state occupation probability in the full- (red solid lines) and the effective-Hamiltonian picture (blue dashed lines). The qubits, the nanoantenna and the driving field are assumed to be on resonance. The system is initially set to its ground state. The following normalized parameters were chosen: (a) weak-coupling regime: $g=0.01\Gamma$, $\Omega=0.05\Gamma$; (b) intermediate case: $g=0.1\Gamma$, $\Omega=0.5\Gamma$; (c) strong-coupling regime: $g=\Gamma$, $\Omega=0.5\Gamma$. In each case $\gamma = 10^{-8}\Gamma$.}
\end{figure}

In Fig.~\ref{fig:adiabatic_approx} the evolution of the symmetric-state occupation probability, calculated  by the full- and effective-Hamiltonian approaches, i.e. by solving Eq.~(\ref{eq:lindblad}) and Eq.~(\ref{eq:lindblad_reduced}. respectively, are compared. The qubits were chosen to be identical ($\gamma^{(1)}=\gamma^{(2)}\equiv \gamma=10^{-8}\Gamma$), resonant with both the nanoantenna and the driving field ($\omega^{(1)}=\omega^{(2)}=\omega_\mathrm{na}=\omega_\mathrm{dr}$), and symmetrically positioned with respect to the nanoantenna (implying $g^{(1)}=g^{(2)}\equiv g$). The evolution in the full and the effective picture is compared for three values of the normalized coupling strength: $g/\Gamma = 0.01$, $0.1$, $1$, corresponding to the so-called weak qubit-to-nanoantenna coupling regime, an intermediate case, and the strong-coupling regime \cite{Huemmer2013,Artuso2013,Esteban2014}. In each case, a relatively strong driving field was chosen: $\Omega/\Gamma = 0.05$ for the weak-coupling case, and $0.5$ otherwise.

As expected, the effective approach is in perfect agreement with the exact one in the first case [Fig.~\ref{fig:adiabatic_approx}(a)], where the loss rate of the field exceeds the coupling to other evolution channels: $\Gamma\gg |g|,|\Omega|$. Only in such parameter regime the evolution of the field is truly adiabatic, i.e. the annihilation operator can be represented by the quasi-stationary solution given by Eq.~(\ref{eq:a_adiabatic}). As the conditions deviate from the genuine weak-coupling regime, the effective approach becomes approximative [Fig.~\ref{fig:adiabatic_approx}(b)] and eventually gives wrong results [Fig.~\ref{fig:adiabatic_approx}(c)]. However, its capacity lies not only in the fact that it simplifies calculations under the weak-coupling conditions. The main advantage of the effective Hamiltonian approach is that it provides us with an intuitive insight into the processes that are responsible for the evolution of the system, in particular in entanglement-generation mechanisms.

This will become more obvious if the effective Hamiltonian and Lindblad term are transformed to a basis more suitable to describe an effectively hybridized system of two qubits. Such transition will be performed in the following subsection. It permits to analyze the entanglement-generation mechanisms, i.e. to use directly an analytical expression for the definition of conditions, where large degrees of stationary entanglement are expected. Later in Sect. \ref{sec:results}, we will show that the results come up to the expectations beyond the weak-coupling regime too.
\begin{figure}
\includegraphics{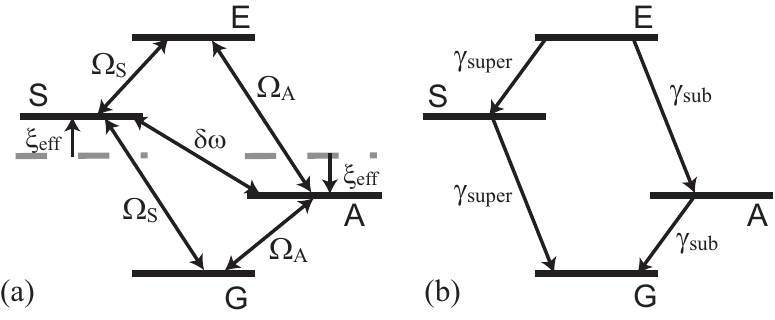}
\caption {\label{fig:hybridization} Scheme of the hybridized two-qubit system represented in the Dicke basis: (a) energy shifts and coherent couplings described by the effective Hamiltonian (\ref{eq:eff_Hamiltonian_Dicke}) and (b) superradiant and subradiant dissipation channels: $\gamma_\mathrm{super}=\left(\gamma_\mathrm{eff}^{(1)}+2\gamma_\mathrm{eff}^\prime+\gamma_\mathrm{eff}^{(2)}\right)/2$, $\gamma_\mathrm{sub}=\left(\gamma_\mathrm{eff}^{(1)}-2\gamma_\mathrm{eff}^\prime+\gamma_\mathrm{eff}^{(2)}\right)/2$, as described in Eqs.~(\ref{eq:dicke3}).}
\end{figure}

\subsection{Transition to the Dicke basis \label{subsec:Dicke}}
The effective Hamiltonian acquires an interesting form in the so-called Dicke basis: $\left\{|\mathrm{E}\rangle = |e^{(1)}e^{(2)}\rangle\right.$, $|\mathrm{S}\rangle = \left(|e^{(1)}g^{(2)}\rangle+|g^{(1)}e^{(2)}\rangle\right)/\sqrt{2}$, $|\mathrm{A}\rangle = \left(|e^{(1)}g^{(2)}\rangle-|g^{(1)}e^{(2)}\rangle\right)/\sqrt{2}$, $\left.|\mathrm{G}\rangle = |g^{(1)}g^{(2)}\rangle \right\}$, where the symbols $\mathrm{E}$ and $\mathrm{G}$ stand for the bi-excited and total ground states, respectively, and $\mathrm{S}$ and $\mathrm{A}$ - for the symmetric and antisymmetric superpositions of states with a single excitation shared between the two qubits \cite{Dicke1954}.
In the Dicke basis the effective Hamiltonian reads
\begin{eqnarray}
H_{\mathrm{eff}}&=&\sum_{m=\mathrm{E,S,A,G}} E_m |m\rangle \langle m|\label{eq:eff_Hamiltonian_Dicke} \\
&&+\delta\omega\left(|\mathrm{A}\rangle\langle \mathrm{S}|+|\mathrm{S}\rangle\langle \mathrm{A}|\right) \nonumber\\
&&-\Omega_\mathrm{S}\left(|\mathrm{S}\rangle\langle \mathrm{G}|+|\mathrm{E}\rangle\langle \mathrm{S}|\right)-\mathrm{H.c.}\nonumber \\
&&-\Omega_\mathrm{A}\left(|\mathrm{A}\rangle\langle \mathrm{G}|-|\mathrm{E}\rangle\langle \mathrm{A}|\right)-\mathrm{H.c.},\nonumber
\end{eqnarray}
In these expressions the physically most important features are (see also Fig.~\ref{fig:hybridization}a):
\begin{itemize}
\item {energy shifts by the effective dipole-dipole term $\xi_\mathrm{eff}$: $E_\mathrm{E}=\Delta\omega^{(1)}_\mathrm{eff}+\Delta\omega^{(2)}_\mathrm{eff}$, $E_\mathrm{S}=\frac{1}{2}(\Delta\omega^{(1)}_\mathrm{eff}+\Delta\omega^{(2)}_\mathrm{eff})+\xi_\mathrm{eff}$, $E_\mathrm{A}=\frac{1}{2}(\Delta\omega^{(1)}_\mathrm{eff}+\Delta\omega^{(2)}_\mathrm{eff})-\xi_\mathrm{eff}$, $E_\mathrm{G}=0$;}
\item {the coherent coupling between the symmetric and antisymmetric states, present if only the effective transition frequencies of the two qubits are shifted with respect to each other: $\delta\omega = \frac{1}{2}(\Delta\omega^{(1)}_{\mathrm{eff}}-\Delta\omega^{(2)}_{\mathrm{eff}})$; as follows from the second of Eqs.~(\ref{eq:eff_param_single_qubit}), such coherent coupling is possible either for non-symmetric positions of the qubits with respect to the nanoantenna, that lead to different coupling strengths $g^{(1)}\neq g^{(2)}$, or for different transition frequencies $\omega^{(1)}\neq \omega^{(2)}$;}
\item {the possibility of a direct access from the ground (and excited) state to the symmetric and antisymmetric states by a coherent coupling to the external field $\Omega$; this coupling is described by the third of Eqs.~(\ref{eq:eff_param_single_qubit}) and effective parameters $\Omega_\mathrm{S} = \frac{1}{\sqrt{2}}(\Omega^{(1)}_\mathrm{eff}+\Omega^{(2)}_\mathrm{eff})$, $\Omega_\mathrm{A} = \frac{1}{\sqrt{2}}(\Omega^{(1)}_\mathrm{eff}-\Omega^{(2)}_\mathrm{eff})$.}
\end{itemize}

The effective Lindblad term (\ref{eq:eff_lindblad}) rewritten in the Dicke basis is somewhat cumbersome, and, therefore, we will now only analyze its contributions to the diagonal terms of the density matrix. They read as:
\begin{eqnarray}
\left[\mathcal{L}_\mathrm{eff}\left(\rho^\mathrm{qb}\right)\right]_\mathrm{EE} &=& -\rho^\mathrm{qb}_\mathrm{EE}\left(\gamma_\mathrm{super}+\gamma_\mathrm{sub}\right), \nonumber\\
\left[\mathcal{L}_\mathrm{eff}\left(\rho^\mathrm{qb}\right)\right]_\mathrm{SS} &=& -\left(\rho^\mathrm{qb}_\mathrm{SS}-\rho^\mathrm{qb}_\mathrm{EE}\right)\gamma_\mathrm{super}\nonumber \\
&&+\frac{1}{2}\Re\left(\rho^\mathrm{qb}_\mathrm{AS}\right)\left(\gamma_\mathrm{eff}^{(1)}-\gamma_\mathrm{eff}^{(2)}\right), \nonumber \\
\left[\mathcal{L}_\mathrm{eff}\left(\rho^\mathrm{qb}\right)\right]_\mathrm{AA} &=& -\left(\rho^\mathrm{qb}_\mathrm{AA}-\rho^\mathrm{qb}_\mathrm{EE}\right)\gamma_\mathrm{sub}\nonumber \\
&&+\frac{1}{2}\Re\left(\rho^\mathrm{qb}_\mathrm{AS}\right)\left(\gamma_\mathrm{eff}^{(1)}-\gamma_\mathrm{eff}^{(2)}\right), \label{eq:dicke3} \\
 \left[\mathcal{L}_\mathrm{eff}\left(\rho^\mathrm{qb}\right)\right]_\mathrm{GG} &=& \rho^\mathrm{qb}_\mathrm{SS}\gamma_\mathrm{super}+\rho^\mathrm{qb}_\mathrm{AA}\gamma_\mathrm{sub}\nonumber\\
 &&-\Re\left(\rho^\mathrm{qb}_\mathrm{AS}\right)\left(\gamma_\mathrm{eff}^{(1)}-\gamma_\mathrm{eff}^{(2)}\right).\nonumber
\end{eqnarray}
Thus there are two decay channels in the system (see Fig.~\ref{fig:hybridization}b): a superradiant one from the bi-excited, via the symmetric, to the ground state, with the rate $\gamma_\mathrm{super}\equiv\left(\gamma_\mathrm{eff}^{(1)}+2\gamma_\mathrm{eff}^\prime+\gamma_\mathrm{eff}^{(2)}\right)/2$; and a subradiant one, via the antisymmetric state, with the rate $\gamma_\mathrm{sub}\equiv\left(\gamma_\mathrm{eff}^{(1)}-2\gamma_\mathrm{eff}^\prime+\gamma_\mathrm{eff}^{(2)}\right)/2$. The population dynamics is additionally complicated by a coupling to the coherence $\rho^\mathrm{qb}_\mathrm{AS}$.

The interpretation becomes more straight while considering identical qubits placed in highly symmetric positions with respect to the antenna, i.e. for $\omega^{(1)}=\omega^{(2)}$, $g^{(1)}=g^{(2)}$ and $\gamma^{(1)}=\gamma^{(2)}$. Then, without driving field $\Omega$, the Hamiltonian is diagonal in the Dicke basis, as can be clearly seen from Eqs.~(\ref{eq:eff_param_single_qubit},\ref{eq:eff_param_collective}) and (\ref{eq:eff_Hamiltonian_Dicke}). Thus, the effective evolution of the Dicke-states occupation probabilities is only given by the Lindblad term (\ref{eq:dicke3}): $\dot{\rho^\mathrm{qb}}_{mm}=\left[\mathcal{L}_\mathrm{eff}\left(\rho^\mathrm{qb}\right)\right]_{mm}$. Moreover, $\gamma^{(1)}_\mathrm{eff}=\gamma^{(2)}_\mathrm{eff}\equiv\gamma_\mathrm{eff}$, and therefore the superradiant rate simplifies to $\gamma_\mathrm{eff}+\gamma_\mathrm{eff}^\prime$, and the subradiant rate is equal to the vacuum-induced spontaneous decay rate $\gamma^{(1)}=\gamma^{(2)}\equiv\gamma$ \cite{Dicke1954,Dzsotjan2011}. To a good approximation the antisymmetric state is decoupled from the rest of the system at time scales comparable to $1/\gamma_\mathrm{eff}$, but its population decays exponentially at time scales of the free-space spontaneous emission rate $1/\gamma$.

It should be noted, that if the qubits are separated by distances less than $2\pi c/\omega^{({j})}$, coupling to the photonic vacuum itself is a source of sub- and superradiance \cite{Dicke1954,Ficek2002}. Then, the lifetime of the antisymmetric state is even longer, which naturally promotes the preservation of entanglement (see also Ref.~\onlinecite{Lidar1998}). In this paper we disregard this simple effect and prove that even when the antisymmetric state is not entirely decay-free, large values of stationary entanglement can be reached.

\subsection{Evaluation of the effective formalism in the context of entanglement generation \label{subsec:discussion}}
Writing down the expression for concurrence one can obtain in the effective picture for a weak driving field $\Omega$. In this case, the coherence parameters $\rho^\mathrm{qb}_{\mathrm{E}p}$, $p\in \{A,S,G\}$ are small and the concurrence reads \cite{Zubairy2011}
\begin{eqnarray}
C_\mathrm{eff}(\rho^\mathrm{qb}) &\approx& \mathrm{max}\left\{0,C_\mathrm{M}\right\}, \label{eq:concurrence_reduced}\\
C_\mathrm{M} &=& \sqrt{\left(\rho^\mathrm{qb}_\mathrm{SS}-\rho^\mathrm{qb}_\mathrm{AA}\right)^2+4\Im\left(\rho^\mathrm{qb}_\mathrm{SA}\right)^2}-2\sqrt{\rho^\mathrm{qb}_\mathrm{GG}\rho^\mathrm{qb}_\mathrm{EE}}.\nonumber
\end{eqnarray}

Having Eqs.~(\ref{eq:eff_Hamiltonian_Dicke}-\ref{eq:concurrence_reduced}) at hand, we can summarize the effective picture obtained in this section from the entanglement-generation point of view:
\begin{itemize}
\item {Eq.~(\ref{eq:concurrence_reduced}) suggests that in the investigated configuration, the entanglement can attain large values when the qubit-qubit system is driven into either the symmetric or the antisymmetric state with a high probability, and at the same time the probability of the bi-excited state $|E\rangle$ remains low.}
\item {In this context, the configuration with equal qubit-to-nanoantenna coupling strengths $g^{(1)}=g^{(2)}$ is especially interesting. In this case, the existence of an approximately nondecaying antisymmetric state, which is at the same time a maximally entangled state, follows from Eqs.~(\ref{eq:dicke3}). This opens an opportunity to produce high degrees of entanglement between the qubits being robust against dissipation. The other maximally entangled state, the symmetric one, decays quickly due to superradiance.}
\item {It turns out, however, as can be seen from Eq.~(\ref{eq:eff_Hamiltonian_Dicke}), that it is the symmetric state that can be directly accessed by a coherent drive (for $g^{(1)}=g^{(2)}$ the effective coupling to the antisymmetric state disappears naturally: $\Omega_\mathrm{A}=0$). The antisymmetric state is decoupled from the rest of the Hilbert space, unless a certain degree of asymmetry between the qubits is introduced, resulting in $\delta\omega\neq 0$. Because we have already set $g^{(1)}=g^{(2)}$, the asymmetry must be evoked by different transition frequencies $\omega^{(1)}\neq \omega^{(2)}$. Then, a high occupation probability of the (approximately) nondecaying antisymmetric state can be reached through a coherent coupling $\delta\omega$ to the symmetric one.}
\end{itemize}
We conclude that two qubits symmetrically coupled to a nanoantenna mode should be efficiently driven into the antisymmetric state if their transition frequencies are different from each other. Then, the population can be driven from the ground to the symmetric state with a coherent drive $\Omega_\mathrm{S}$, and from the symmetric state to the antisymmetric one, by the coupling induced by the difference in the transition frequencies $\delta\omega$. The small decay rate of the antisymmetric state allows it to remain highly-occupied, which leads to a large concurrence. This scheme is at the heart of our proposal.

In this section we have developed a simplified approach to describe the qubit-qubit dynamics for qubits weakly coupled to a lossy nanoantenna field mode. By an adiabatic elimination of the operators associated with the field, we arrived at an effective Lindblad-Kossakowski formalism. Even though the effective picture is only valid  in the weak qubit-to-nanoantenna coupling regime, it is simple and instructive. Moreover, as we will show in Sect. \ref{sec:results}, its predictions for entanglement-generation mechanisms hold beyond this regime. As the conclusion of this section we find that two qubits symmetrically coupled to a nanoantenna mode should be efficiently driven into the antisymmetric state if only their transition frequencies differ from each other.

\section{Asymmetry-induced interqubit entanglement\label{sec:results}}
In this section we verify the insights gained above. First, we solve the Lindblad-Kossakowski equation (\ref{eq:lindblad}) with the full Hamiltonian (\ref{eq:hamiltonian}) for the perfect case of two qubits characterized by equal coupling strengths to the nanoantenna mode: $g^{(1)}=g^{(2)}\equiv g$. The transition frequencies of the qubits are anti-symmetrically detuned from the nanoantenna resonance: $\omega^{(1)} = \omega_\mathrm{na}+\delta\omega$, $\omega^{(2)}=\omega_\mathrm{na}-\delta\omega$.

\subsection{Perfect case of equal coupling strengths and anti-symmetric detuning}
\begin{figure}
\includegraphics{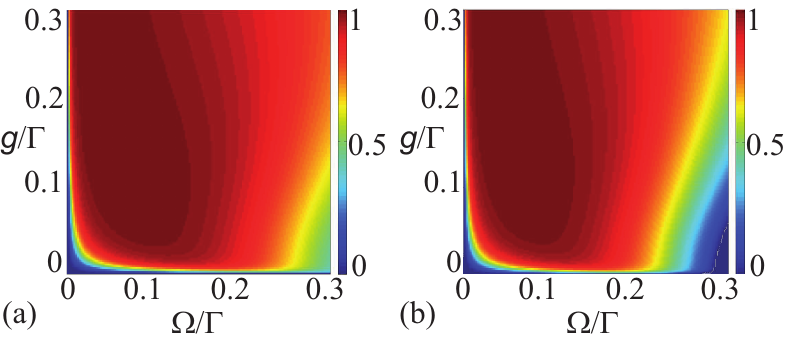}
\caption {\label{fig:results_g_Omega} (a) The steady-state probability of the antisymmetric state $\rho_\mathrm{AA}(t\rightarrow\infty)$ and (b) the steady-state concurrence $C$ as functions of the normalized coupling constant $g/\Gamma$ and driving field $\Omega/\Gamma$, for two qubits anti-symmetrically detuned from the nanoantenna resonance: $\delta\omega = \omega^{(1)}-\omega_\mathrm{na} = \omega_\mathrm{na}-\omega^{(2)} = 10^{-3}\Gamma$, and vacuum-induced spontaneous emission rate $\gamma=10^{-8}\Gamma$. }
\end{figure}

In Fig.~\ref{fig:results_g_Omega}, steady-state values of the antisymmetric-state probability $\rho_\mathrm{AA}(t\rightarrow\infty)$ and the corresponding concurrence $C$, are plotted as functions of the normalized coupling strength $g/\Gamma$ and driving-field Rabi frequency $\Omega/\Gamma$, for the detuning $\delta\omega=10^{-3}\Gamma$ and the vacuum-induced spontaneous-emission rate $\gamma^{(1)}=\gamma^{(2)}\equiv\gamma=10^{-8}\Gamma$. The parameter range analyzed in Fig.~\ref{fig:results_g_Omega} corresponds to nanoantennas investigated in Ref.~\onlinecite{Slowik2013}: silver nanospheres and nanospheroids of both the characteristic size and the qubit-to-nanoantenna distance up to $100$ nm.

As expected, it follows from Fig.~\ref{fig:results_g_Omega} that the obtained stationary entanglement can be extremely large. Naturally, both $\rho_\mathrm{AA}(t\rightarrow\infty)$ and $C$ are equal to zero for a vanishing coupling to the nanoantenna resonance ($g=0$) or without driving field ($\Omega=0$). The results quickly grow with $g$ and small values of $\Omega$ and reach unity or almost unity for coupling constants $g>0.05\Gamma$, for a considerable range of $\Omega$s. As follows from Eq.~(\ref{eq:eff_Hamiltonian_Dicke}), for strong driving fields the role of the processes that distribute the population between the $|G\rangle$, $|S\rangle$ and $|E\rangle$ states (the terms proportional to $\Omega_\mathrm{S}$) is increased with respect to the coherent transfer between the symmetric and antisymmetric states (the term $\sim\delta\omega$). This is why $\rho_\mathrm{AA}(t\rightarrow\infty)$ drops for large values of $\Omega$. The concurrence is further decreased due to the increased probability of the bi-excited state occupation $\rho_\mathrm{EE}(t\rightarrow\infty)$ [see the second term in Eq.~(\ref{eq:concurrence_reduced})]. However, it is important to stress that a concurrence close to unity can be obtained in a large parameter space.

\subsection{Analyzing the robustness of the scheme}
\begin{figure}
\includegraphics{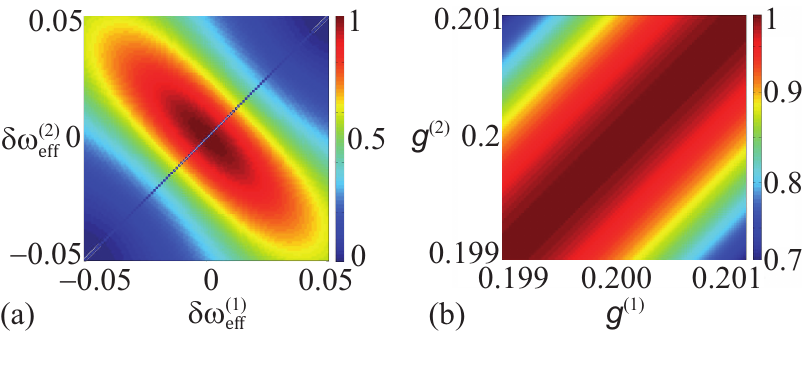}
\caption {\label{fig:how_robust} (a) The steady-state concurrence $C$ as a function of normalized detunings of the qubits $\delta\omega^{(1)}/\Gamma$ and $\delta\omega^{(2)}/\Gamma$, for the coupling constant $g=0.2\Gamma$. (b) Concurrence as a function of different qubit-nanoantenna coupling constants $g^{(1)}$ and $g^{(2)}$, for a symmetric detuning $\delta\omega = 10^{-3}\Gamma$. In both cases the driving field $\Omega=0.1\Gamma$, vacuum-induced spontaneous emission rate $\gamma=10^{-8}\Gamma$.}
\end{figure}

It is important to double-check the sensitivity of the results presented in Fig.~\ref{fig:results_g_Omega} against variations of nominal parameters of the system which might be eventually caused by experimental imperfections. The steady-state concurrence as a function of the normalized qubit-transition-frequency detunings from the nanoantenna resonance $\delta\omega^{({j})}/\Gamma \equiv \left(\omega^{({j})}-\omega_\mathrm{na}\right)/\Gamma$ is shown in Fig.~\ref{fig:how_robust}(a) for $g = \Omega =0.2\Gamma$, $\gamma = 10^{-8}\Gamma$. As expected, the concurrence vanishes for identical qubits, i.e. for $\delta\omega^{(1)}=\delta\omega^{(2)}$, and grows fast for very small transition-frequency differences (the apparent discontinuity is a result of imperfect resolution of the figure). The result proved to be robust against small deviations from the optimal conditions: the concurrence remains high even if the absolute values of the detunings $\delta\omega^{({j})}$ are not equal, as long as they have opposite signs. (For the nanoantenna parameters described in Ref.~\onlinecite{Slowik2013}, this corresponds to a tolerance of a few THz for the concurrence to exceed $0.9$.) When the detunings are too large, however, the qubits are no longer in exact resonance with the nanoantenna and the driving field, which leads to smaller concurrence.

We proceed in analyzing the steady-state concurrence for unequal coupling constants $g^{(1)}\neq g^{(2)}$. The result is shown in Fig.~\ref{fig:how_robust}(b) for $\delta\omega=10^{-3}\Gamma$, $\Omega=0.1\Gamma$, $\gamma = 10^{-8}\Gamma$. Interestingly, even though such imperfect configuration leads to additional asymmetry in the system, it results in smaller degrees of steady-state entanglement. This is because for unequal coupling constants the antisymmetric state undergoes a decay with a rate proportional to $\left(g^{(1)}-g^{(2)}\right)^2$ [see the third of Eqs.~(\ref{eq:dicke3})]. Consequently, its steady-state probability is smaller. This is in agreement with the results obtained in Refs.~\onlinecite{MartinCano2011,Chen2012}, where asymmetric placements of the qubits with respect to a nanowire or a microtoroid were considered for the purpose of entanglement generation at short time-scales. The resulting entanglement did not, however, persist in the steady-state. Such sensitivity of the concurrence to the coupling strengths implies in particular a need for precisely placing the qubits with respect to the nanoantenna surface. This seems to be within the reach of the state-of-the-art technology, which allows for a control over the qubit-to-nanoantenna distance with a sub-nanometer resolution \cite{Benson2009,Hartschuh2011,Alaee2013,Gruber2013,Belacel2013,Giessen2014}.

In the last section we have numerically solved the Lindblad-Kossakowski equation to confirm that a very large degree of stationary interqubit entanglement, with concurrence reaching one, can be obtained with two qubits of different transition frequencies, symmetrically coupled to a nanoantenna. We have found that the scheme is robust against perturbations in the detuning of the qubits from the nanoantenna resonance. For the steady-state concurrence to be large it is however crucial that the antisymmetric state remains strongly subradiant, i.e. that the effective loss rates of all qubits are equal. This requires a precise positioning of the qubits with respect to the nanoantenna.

\section{Conclusions\label{sec:conclusions}}
A cavity-QED formulation of the problem of two qubits coupled to a single nanoantenna resonance has been considered in the context of interqubit entanglement generation.

The underlying physical mechanisms were analyzed by a simplistic effective description, strictly valid in the weak-coupling regime only. The effective description directly leads to the conclusion that the qubits may be driven into a maximally entangled steady state for symmetric coupling to the mode of the nanoantenna, but an antisymmetry in the transition frequencies of the qubits.

Great advantages of the proposed scheme are that (1) it is independent of the initial state of the system; and (2) it is robust against dissipation through channels associated with both the nanoantenna and the qubits. However sensitive to deviations in the coupling strength between the qubits and the nanoantenna field, the proposed scheme has been proven stable with respect to perturbations in their spectral properties.

Implementation of entanglement generation and more complicated quantum-information protocols with plasmonic structures promises faster quantum computing with miniaturized devices.

\section*{Acknowledgements}
This work was partially supported by the German Federal Ministry of Education and Research (PhoNa) and by the Thuringian State Government (MeMa).

\end{document}